\documentclass[final]{agujournal2019}
\usepackage{natbib, graphicx}
\usepackage{url} 
\usepackage{lineno}
\usepackage[inline]{trackchanges} 
\usepackage{soul}

\soulregister\ref7
\soulregister\cite7
\soulregister\citeA7
\soulregister\citep7
\soulregister\subsection7

\usepackage[version=4]{mhchem}
\graphicspath{{/Users/eyigit/OneDrive/Research/Data/GOLD-ICON/Figures/}{./Figures}}
\drafttrue

\usepackage{bm} 
\usepackage{microtype}
\newcommand{\del}{\bm{\nabla}}

\newcommand{\Figures}{\textit{Figures}~}
\newcommand{\Figure}{\textit{Figure}~}

\definecolor{citegrey}{rgb}{0.4, 0.5, 0.6}
\definecolor{darkred}{rgb}{0.65, 0.25, 0.2}
\definecolor{citeE}{rgb}{0.1, 0.5, 0.9}

\journalname{Journal of Geophysical Research -  Space Physics}

\usepackage{marginnote}

\newcommand{\laplacian}{{\bm\nabla}^2}

\begin{document}
\title{Characterization of the Thermospheric Mean Winds and Circulation during Solstice using ICON/MIGHTI Observations}

\authors{Erdal Yi\u git\affil{1}, Manbharat Dhadly\affil{2}, Alexander S. Medvedev\affil{3}, Brian J. Harding\affil{4}, Christoph R. Englert\affil{2}, Qian Wu\affil{5,6}, Thomas J. Immel\affil{4}}
\affiliation{1}{George Mason University, Department of Physics and Astronomy, Space Weather Lab,  Fairfax, VA, USA.}
\affiliation{2}{U.S. Naval Research Laboratory, Washington D.C., USA.}
\affiliation{3}{Max Planck Institute for Solar System Research, G\"ottingen, Germany}
\affiliation{4}{UC Berkeley, Space Sciences Laboratory, Berkeley, CA, USA.}
\affiliation{5}{High Altitude Observatory, NCAR, Boulder, CO, USA}
\affiliation{6}{COSMIC Program UCAR/UCP, Boulder, CO, USA}

\correspondingauthor{Erdal Yi\u git}{eyigit@gmu.edu}

\begin{keypoints}
\item Mean zonal and meridional winds are derived for Northern summer and winter solstice conditions from ICON/MIGHTI observations
\item Horizontal winds exhibit a significant degree of spatiotemporal variability,  exceeding $\pm 150 $ m s$^{-1}$.
\item Zonal and meridional mean winds exhibit reversal in the lower thermosphere.
\item Distributions of mean winds and circulation are more homogeneous in the upper thermosphere than lower thermosphere.
\end{keypoints}

\begin{abstract}
Using the horizontal neutral wind observations from the MIGHTI instrument onboard NASA's ICON (Ionospheric Connection Explorer) spacecraft with continuous coverage, we determine the climatology of the mean zonal and meridional winds and the associated mean circulation at low- to middle latitudes ($10^\circ$S--40$^{\circ}$N) for Northern Hemisphere {summer} solstice conditions between 90 km and 200 km altitudes, specifically on 20 June 2020 solstice as well as for a one-month period from 8 June--7 July 2020 {and for Northern winter season from 16 December 2019--31 January 2020, which spans a 47-day period, providing full local time coverage}. The data are averaged within appropriate altitude, longitude, latitude, solar zenith angle, and local time bins to produce mean wind distributions. The geographical distributions and local time variations of the mean horizontal circulation are evaluated. The instantaneous horizontal winds exhibit a significant degree of spatiotemporal variability often exceeding $\pm 150 $ m s$^{-1}$. The daily averaged zonal mean winds demonstrate day-to-day variability. Eastward zonal winds and northward (winter-to-summer) meridional winds are prevalent in the lower thermosphere, which provides indirect observational evidence of the eastward momentum deposition by small-scale gravity waves. The mean neutral winds and circulation exhibit smaller scale structures in the lower thermosphere (90--120 km), while they are more homogeneous in the upper thermosphere, indicating the increasingly dissipative nature of the thermosphere. The mean wind and circulation patterns inferred from ICON/MIGHTI measurements can be used to constrain and validate general circulation models, as well as input for numerical wave models.
\end{abstract}

\section*{Plain Language Summary}
Atmospheric horizontal winds (i.e., motion of the neutral air), composed of zonal (east-west) and meridional (north-south) components, play an important role for the energy and momentum balance of the atmosphere and ionosphere. Due primarily to a lack of observations, winds in the thermosphere are not well sampled. In this study we use the horizontal winds measured from 90--200 km altitude by the MIGHTI instrument onboard NASA's ICON (Ionospheric Connection Explorer) spacecraft  to generate two-dimensional maps of zonal and meridional winds, and of the resulting horizontal motion (or circulation) in the thermosphere for Northern Hemisphere solstice conditions. Specifically, winds at solstice (20 June 2020) and a one-month {Northern summer} solstitial period (8 June--7 July 2020) {and a 47-day winter solstitial period (16  December 2019--31 January 2020)} have been analyzed. Mean winds show significant spatial variation as a function of time, often demonstrating tidal variability.

\section{Introduction}
Earth's thermosphere extending from $\sim $90 km upwards is the outermost region of the atmosphere, where satellites orbit the planet and a substantial portion of solar ultraviolet (UV) radiation is absorbed by atmospheric gases. This rarefied and highly dissipative region is influenced by a broad spectrum of internal atmospheric waves propagating upward from the  lower atmosphere \citep{Hickey_etal11, PanchevaMukhtarov12, YigitMedvedev15, Oberheide_etal15, Gavrilov_etal20, Pancheva_etal20, Forbes_etal21, Dhadly2018b} and by solar and geomagnetic processes (i.e., space weather)  from above \citep{SchunkSojka96, Emmert15, Yigit_etal16b, Deng_etal18, Ward_etal21, ShiokawaGeorgieva21, Dhadly2018}. The forces acting on the neutral flow are rarely in balance in the thermosphere, thus giving rise to an enhanced spatiotemporal variability with turbulent to global scales.

The goal of this paper is to characterize the thermospheric mean zonal and meridional winds, and circulation during solstice at low- to midlatitudes (10$^\circ $S--40$^\circ$N) using observations from the MIGHTI (Michelson Interferometer for Global High-resolution Thermospheric Imaging) instrument \citep{Englert_etal17}  onboard NASA's ICON (Ionospheric Connection Explorer) spacecraft during 2020 Northern Hemisphere solstice conditions.  Due primarily to poor observational coverage, neutral winds have been insufficiently characterized at thermospheric altitudes. Horizontal winds and the associated circulation  play an essential role in the energy and momentum budget of the thermosphere and ionosphere. They modulate upwelling/downwelling of air, influence the critical filtering and dissipation of internal atmospheric waves propagating upward, drive low-latitude ionospheric electrodynamics, transport major chemical species (e.g., \ce{O}, \ce{N2}, \ce{NO}), generate neutral drag on the ions, and redistribute thermospheric energy and momentum in general. Thermospheric winds are primarily horizontal, however, in the regions of convergence or divergence, upwelling or downwelling (i.e., vertical motion) can occur as a consequence of the principal of conservation of mass \citep{Rishbeth_etal69, Smith98}. Therefore, characterization of the mean winds is essential for our understanding of the thermosphere-ionosphere system as a whole.

Various methods were utilized to observe winds over a broad range of altitudes from the upper mesosphere to the thermosphere. However, lower thermospheric winds are more routinely observed than in the upper thermosphere. A summary of historical and current observations is presented by other researchers \citep[e.g., ][]{Drob_etal08, Drob_etal15,Dhadly2019}. Chemical release wind measurements carried out in different sites around the world can provide profiles of wind velocity and {high vertical resolution} wind shear from $\sim$80 -- 140 km \citep{Larsen02, Lehrmacher_etal22}. Incoherent scatter radars around the world use measured ion drifts to derive neutral winds from $\sim90 - 130$ km \citep{Zhang_etal03, Hysell_etal14}. Meteor echoes and meteor radars are used to retrieve wind profiles between $\sim90 - 110$ km \citep{Oppenheim_etal09, Conte_etal22}. Various types of ground-based Fabry-Perot Interferometers (FPI)  have been employed since {the} 1980s across the globe to understand neutral wind dynamics of the upper and lower thermospheric winds \citep{Conde1995, Meriwether2006, Aruliah2010, Makela2012}.

Satellites can provide measurements of thermospheric winds at a broad range of thermospheric altitudes. Dynamic Explorer 2 (DE2), was the first to monitor upper thermospheric neutral winds from space utilizing a FPI \citep{Killeen1988}. The wind imaging interferometer (WINDII) aboard the Upper Atmosphere Research Satellite (UARS) retrieved neutral winds based on the interfermetric limb measurements of the visible airglow emissions of  {557.7} nm \ce{O^1S} (green line) and 630.0 nm \ce{O^1D} (redline)  between 90 and 300 km \citep{Emmert_etal01, Shepherd_etal12}.  Thermosphere, Ionosphere, Mesosphere Energetics and Dynamics/TIMED Doppler Interferometer (TIMED/TIDI) primarily focused on monitoring MLT winds, launched in 2001 is still operational after 20 years in orbit \citep{Niciejewski2006}. Cross-track winds were derived from accelerometer measurements between 250 km and 400 km by the GOCE (Gravity Field and Steady-State Ocean Circulation Explorer \citep{doornbos2014}) and CHAMP (CHAllenging Minisatellite Payload) satellite \citep{LiuH_etal06, Lieberman_etal13}.

Despite the extensive measurements by ground-based and space-borne instruments, thermospheric winds have been insufficiently sampled so far. Much of the understanding of the thermospheric winds is based on dedicated first-principle global scale modeling \cite[e.g., ][]{Geisler66, Richmond_etal92, Vichare_etal12, Miyoshi_etal14, Yigit_etal16a, Deng_etal18} and empirical models are routinely used to study the global behavior of winds \citep{Drob_etal15, Dhadly2019}. Depending on the type of model and observations, model-data agreement is often partially achieved \citep[e.g.,][]{Tang_etal21}. Thermospheric wind is an essential input parameter for ionospheric models and a better representation of neutral winds is needed for improved space weather modeling \citep{David_etal14}. While models are powerful tools to study the different forces shaping the winds, often the simulated winds are insufficiently constrained in models, therefore ground-based and space-borne measurements of neutral winds are crucial for validating first principal models and to obtain a more complete physical understanding of thermospheric dynamics. 

Although ICON has started observing the thermosphere only recently, MIGHTI observations have already been used to study thermospheric winds and to compare them to other ground-based and space-borne instruments. Recent studies have validated some aspects of ICON horizontal winds with respect to Fabry-Perot interferometers and meteor radars \citep{Harding_etal21, Makela_etal21, Chen_etal22}. \citet{Dhadly_etal21} compared MIGHTI winds to the University of Michigan TIMED Doppler Interferometer (TIDI) level 3 data, contributing to the validation of MIGHTI winds as well as providing guidance towards improving TIDI winds. In addition, this study revealed the longitudinal variations in neutral winds associated with non-migrating tides, which are currently missing from the existing wind climatologies. \citet{Forbes_etal21} analyzed coincident ICON measurements of neutral horizontal winds, ion drifts, and densities and demonstrated a direct link between the day-to-day variability of the wave-4 structure in the E-region and drifts and
densities of ions in the F-region ionosphere.
{
  More recently, \citet{England_etal22}'s analysis of MIGTHI winds have shown that
  strong wind shear is a common feature of the lower thermosphere between 100-130
  km. \citet{Yamazaki_etal22} have analyzed concurrent observations from COSMIC-2 and ICON/MIGHTI wind and found a correlation between the negative vertical shear of the eastward wind and the occurrence rate of the sporadic E layer.
}
Characterization of varying fields is often represented in the form of appropriately defined mean, which requires a sufficient degree of observational coverage in space and time. The averaging for a field variable $\psi$ that is often conducted over time $t$ and longitude $x$ 
\begin{linenomath}
\begin{equation}
  \label{eq:averaging}
  \bar{\psi}(z,t) = \frac{1}{\Delta x\,\Delta \tau}\int_t^{t+\Delta \tau}\int_x^{x+\Delta x} \psi(x,y,z,t) \, dx dt
\end{equation}
\end{linenomath}
is generally referred to as ``zonal mean", if $\Delta x$ spans all longitudes.  In this paper, we perform averaging of ICON/MIGHTI neutral winds over longitude,  latitude and local time, and generally call the results  ``mean winds" as well. Besides the physical importance of winds and their mean structure discussed above, they are routinely used to validate theory and global scale models (or general circulation models) in the middle and upper atmosphere \citep[e.g.,][]{Lieberman_etal00, Garcia_etal07, Dempsey_etal21, Griffith_etal21, Yigit_etal21a, Koval_etal22}.

This paper analyzes the thermospheric horizontal winds between 90  and 200 km during June solstice conditions as observed in 2020 by ICON/MIGHTI . Next section describes the MIGHTI neutral wind measurements used in this study; Section \ref{sec:results} presents the results for the solstice wind distribution and circulation; Section \ref{sec:discussion} provides a discussion of the observed winds, and a summary and conclusions are given in Section \ref{sec:summary-conclusions}.

\section{Materials and Methods}

\subsection{ICON Mission and Data}
\label{sec:icon}
The ICON mission was launched in 2019 and has been surveying the low-latitude thermosphere-ionosphere system above 90 km in unprecedented detail. Its primary goal is to explore Earth's thermosphere-ionosphere system and its connection to geospace as well as terrestrial drivers \citep{Immel_etal18}. The MIGHTI wind observations are based on the Doppler shift measurements the of the green line ($\lambda= 557.7$ nm) and red line ($\lambda = 630$ nm)  emissions of atomic oxygen. In this paper, we analyze the cardinal winds (i.e., zonal  $u$ and meridional  $v$ components) from the MIGHTI instrument \citep{Englert_etal17}. The details of the wind retrieval algorithm are described in the work by \cite{Harding_etal17}. In the following, we outline the data selection, quality, and spatiotemporal coverage.  

In this study, we focus only on the MIGHTI green line neutral winds, which cover the altitude range from $\sim$90 to 200 km during daytime and  $\sim $90 to $\sim $115 km both daytime and nighttime.  Typically, MIGHTI daytime line of sight (LOS) wind observations are available at a 30-s cadence, while nighttime LOS wind measurements are available at a 60-s cadence. Northward and eastward components of the winds are obtained by pairing these LOS wind measurements of MIGHTI A and MIGHTI B sensors, which are taken approximat{e}ly 8 minutes apart.

The quality of MIGHTI/ICON data was taken into account in the analyses. Each  wind measurement was assigned a data quality flag corresponding to  ``good", ``good, but use with caution", and ``bad". We removed all data with bad quality and also excluded outliers with wind magnitudes exceeding 300 m~s$^{-1}$. The result is shown in  \Figure \ref{fig:wind-quality-junjul2020}, where the zonal winds are plotted with different quality flags. The typical accuracy of winds derived from the green line emission is 12 m~s$^{-1}$ or better, as compared with meteor radars \citep{Harding_etal21}. It is seen that such procedure leaves a significant amount of data to maintain solid statistics.

{
  In general, strong mean winds (jets) develop in both hemispheres in the  middle atmosphere during solstices. They produce  enhanced wave filtering and/or modify propagation conditions for internal atmospheric waves with great dynamical implications for the upper atmosphere. Therefore, the solstice winds and circulation are highly variable and interesting to study. Gravity waves and tides significantly affect the mean flow  in the whole atmospheric system. During the Northern winter season additional effects of planetary waves and occurrence of sudden stratospheric
  warming (SSW) events can complicate the interpretation of observations. Therefore,
  we have focused in this study primarily on the Northern  Hemisphere summer solstice (Sections 3.1--3.2). However, in order to complement  June/July solstice results,  we have added a complementary analysis for December 2019--January 2020 Northern winter solstice  (Section 3.3.), which did not include sudden warming events. 
}

\subsection{ICON coverage}
\begin{figure}
  \includegraphics[width=1.1\textwidth, keepaspectratio, height=0.9\textheight]{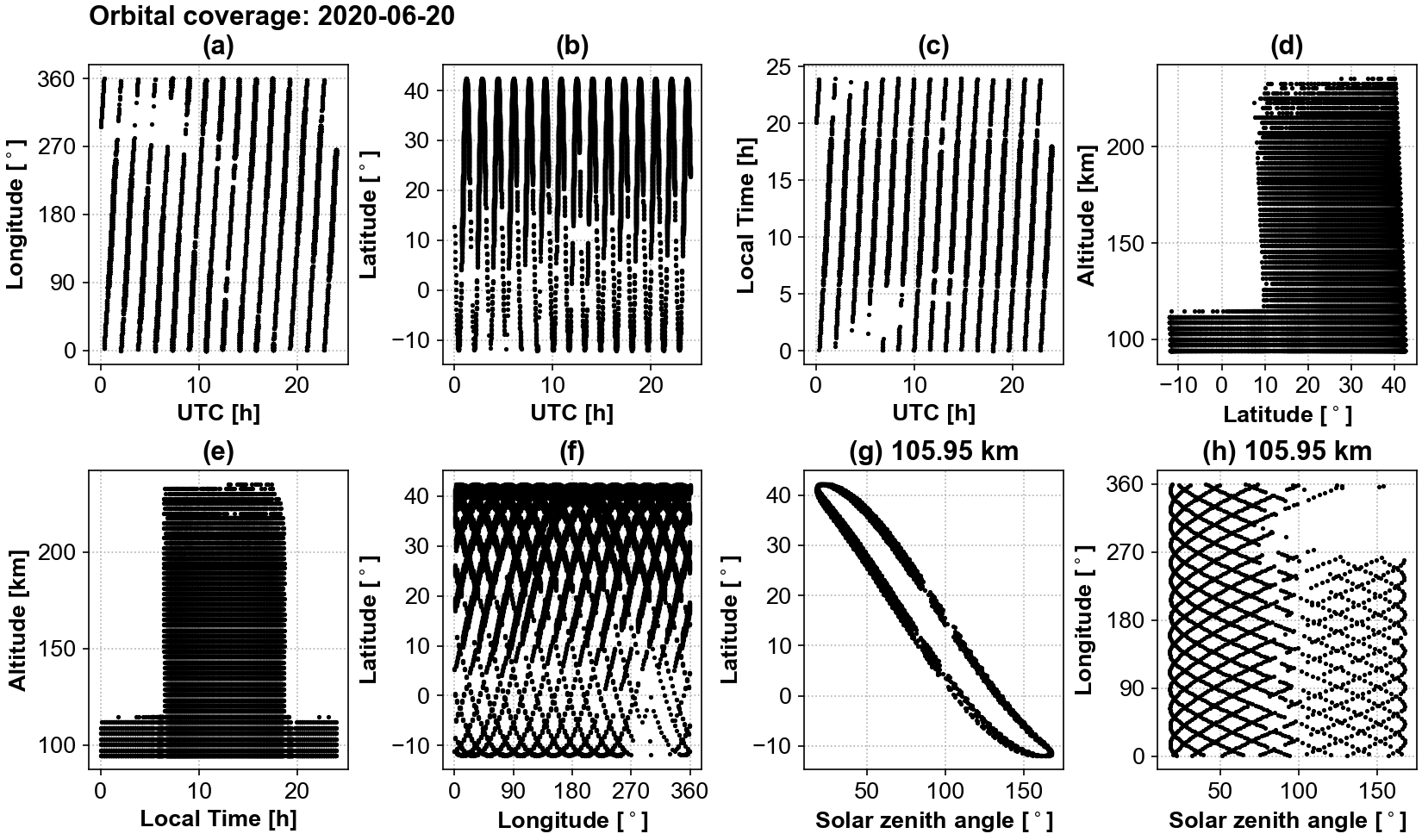}
\caption{Spatiotemporal coverage of ICON/MIGHTI observations during 20 June 2020. Only good quality data (2206 profiles) have been retained.}
  \label{fig:ICON_coverage}
\end{figure}

\Figure \ref{fig:ICON_coverage} illustrates the spatiotemporal ICON data coverage for 20 June 2020, after having removed the bad quality data.  This subset for the analysis contains 2206 individual wind profiles. Panels $a-c$ show the longitude, latitude, and local time coverage as a function of UTC. It is seen that all longitudes and latitudes between --10$^\circ$ and +40$^\circ$ are well covered. There are some nighttime data gaps, however, overall all local times were observed. Panels $d$ and $e$ present the altitude coverage as a function of latitude and local time. It is seen that during daytime, the altitude coverage extends into the upper thermosphere, while the nighttime observations are available only between 90--115 km. Also, latitudes between --10$^\circ$ and +10$^\circ$ are not observed above 115 km on 20 June, since they coincide with nighttime. The latitude-longitude distribution highlights the good spatial coverage between --10$^\circ$ and +40$^\circ$ for all longitudes, with some data gaps around 300$^\circ$ longitude in the Southern Hemisphere associated with South Atlantic Anomaly (SAA). The latitude-solar zenith angle distribution of measurements (panel $g$) at $\sim $106 km suggests that the Northern Hemisphere is covered primarily at daytime, while the Southern Hemisphere latitudes are observed at nighttime. Longitude-solar zenith angle variations at $\sim $106 km (panel h) show that, for a given longitude, both nighttime and daytime data are available except in the region around 300$^\circ$ longitude. 

\begin{figure}[t]
  \includegraphics[width=1.05\textwidth,  keepaspectratio, height=0.9\textheight]{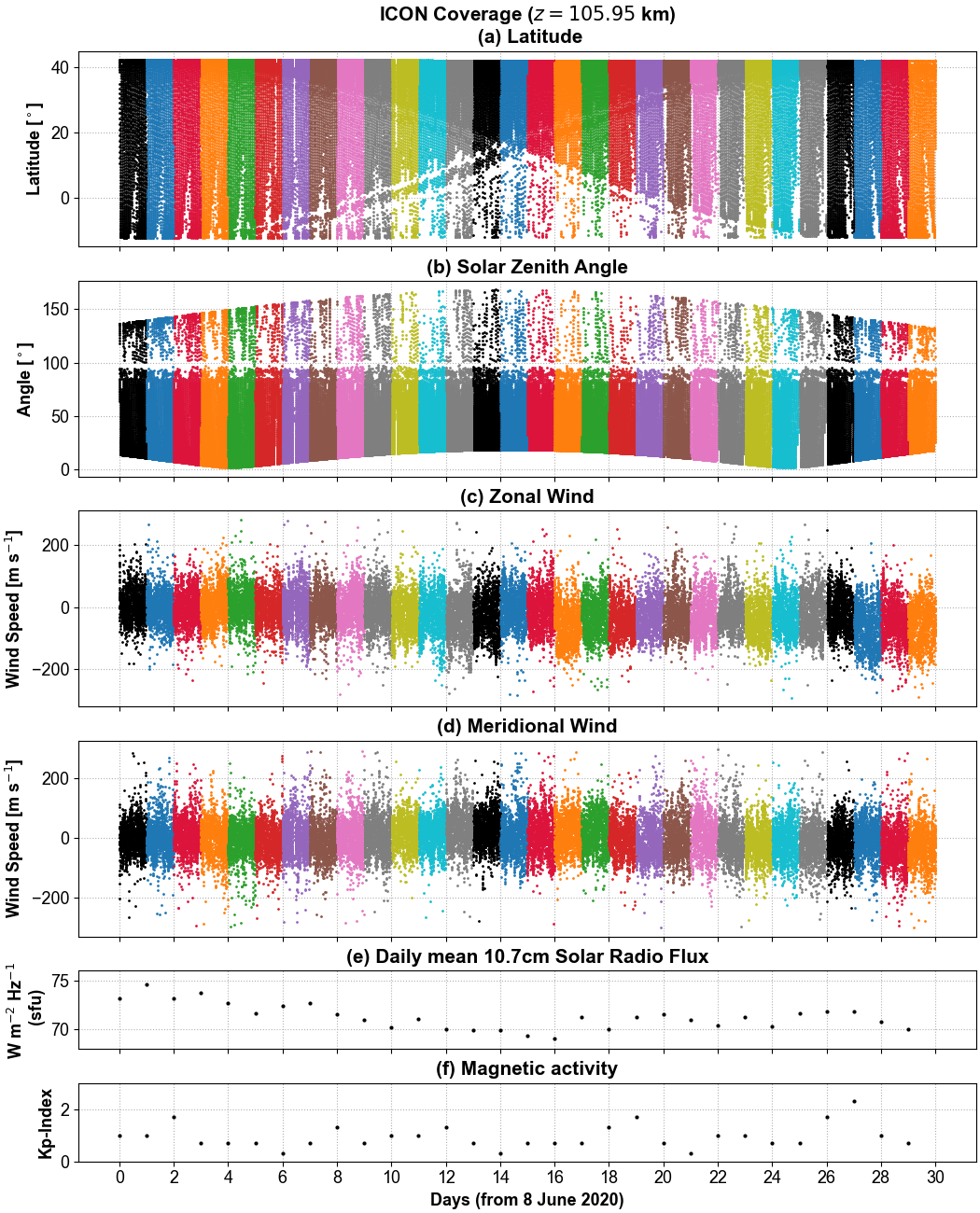}
  \caption{Spatiotemporal coverage of ICON from 8 June to 7 July 2020 (a-b) and winds
    (c-d) at 105.95 km. Daily mean 10.7cm solar radio flux in solar flux units (sfu) is
    shown in panel e {and the geomagnetic activity is shown in terms of the daily
      mean Kp-index in panel f}. The different colors are for the different days.}
  \label{fig:coverage-junjul2020}
  \end{figure}

A single day observation is not sufficient to produce a wind climatology. Therefore, in order to obtain a more consistent picture of mean winds and circulation, we have used one month of continuous ICON observations from 8 June to 7 July 2020, representative of Northern Hemisphere summer solstice conditions. Neutral wind measurements were binned with respect to altitude, latitude, longitude, local time and solar zenith angle with bin sizes of 5 km, 5$^\circ$,
30$^\circ$, 1-2 h, and 10$^\circ$ bins, respectively. \Figure
\ref{fig:coverage-junjul2020} shows the latitude and solar zenith angle coverage along
with the corresponding distributions of the zonal and meridional winds at $\sim$106 km
during this period. {The daily mean F10.7 cm solar radio flux and geomagnetic
  Kp-index}  variations shown in panel{s} $e$ {and
  \textit{f}} indicate quiet solar  {and} {geomagnetic} conditions. The latitudes between 10$^\circ $S and 40$^\circ $N are continuously covered, however, the latter period of June 2020, the southern latitudes are more sparsely sampled than northern latitudes. Also, while all solar zenith angles between 10--140$^{\circ} $ are observed, the nighttime, in general, is observed more sparsely than daytime. The zonal and meridional winds at $\sim$106 km are generally up to $\pm 150$ m~s$^{-1}$ throughout this period, exhibiting noticeable degree of day-to-day variability.     

\section{Results}
\label{sec:results}
\subsection{Zonal and Meridional Winds on 20 June 2020}
\label{sec:mean-winds}

\Figure  \ref{fig:all_winds} shows variations of the zonal (upper panels) and meridional (lower panels) winds with altitude, longitude, latitude, and local time on 20 June 2020. The profiles are plotted without any binning. The average of all vertical profiles are shown with the red line. The zonal and meridional winds demonstrate a high degree of spatiotemporal variability. They are generally faster at higher altitudes, occasionally exceeding $\pm 150$ m~s$^{-1}$. Also, the daytime winds observed in the Northern Hemisphere are faster than those during nighttime at low-latitudes. Note also that the intermittent values of the winds are much larger than their average quantity. Therefore, the latter should be treated with caution as not being representative of instantaneous numbers. 

In order to study the observed wind variability, we have evaluated in \Figure \ref{fig:variability}  the occurrence rates of the wind speeds binned in 5 m~s$^{-1}$ intervals on 20 June 2020 at three representative altitude layers 94--103 km, 106--114 km, and 194--202 km such that each layer included equal number of data points. The speeds shown as a function of number of measurements exhibit a Gaussian distribution generally centered around slow speeds. The associated standard deviations, $\sigma_u$ and $\sigma_v$, which are a proxy for wind variability, are shown in the upper left corner. The nighttime wind variabilities are greater than daytime ones and increase with height. It should be noted that the reported standard deviations are the root mean-square of wind variability and observational errors, which vary with height.

\Figures \ref{fig:zonal_winds_binned} and \ref{fig:meridional_winds_binned} present the altitude distribution of the zonal and meridional winds, respectively, binned as functions of latitude, longitude, local time, and solar zenith angle, where red and blue shadings represent positive (eastward, northward) and negative (westward, southward) wind values, respectively. It is seen that the behavior of the observed zonal and meridional winds in the lower thermosphere (90--110 km) is remarkably different than in the upper thermosphere. Zonal winds are predominantly eastward in the lower thermosphere, in particular, in northern latitudes between 10-40$^\circ $N, with magnitudes of up to 40 m~s$^{-1}$. Above 110 km, they are clearly westward, have speeds exceeding --70 m~s$^{-1}$ and essentially associated with daytime values. The eastward wind regime in the lower thermosphere itself exhibits a substantial degree of variability, when viewed as a function of longitude, local time, and solar zenith angle. For example, the daytime lower thermospheric winds are more eastward compared to those during nighttime. Meridional winds exhibit alternating patterns of flow direction with altitude. They are predominantly southward during daytime in the lower thermosphere between 90--100 km; northward between 100--120 km with speeds reaching 60 m~s$^{-1}$; southward between 120--190 km, and northward again around 190-200 km, especially around noontime, where the winds above 110 km correspond to daytime measurements. The features near the terminator in \Figures 5c,d are potentially an artifact associated with low airglow signal, which are expected to be corrected in {the next} version of the wind data.

\subsection{Thermospheric Mean Winds and Circulation during Northern Hemisphere Summer Solstice}

A diurnal mean provides a short-term glimpse of the circulation, however it is not sufficient for deriving a more accurate view of the climatology of mean winds due to limitations of the orbital coverage to one day and intrinsic variability of the wind field. Therefore, we analyzed one month of continuous ICON observations from 8 June to 7 July 2020, which are representative of the solstitial dynamics during the Northern Hemisphere summer.

\Figure \ref{fig:u_alt_lat_days_junjul} shows the sequence of diurnal-mean altitude-latitude cross-sections of the zonal winds. It clearly shows that the winds strongly vary from day to day. Eastward winds with 10--40 m~s$^{-1}$ speeds are a robust feature of the midlatitude lower thermosphere between 90 and 110 km. The westward winds during daytime dominate above 120 km. Their speeds vary from --10 to --80 m~s$^{-1}$, depending on the latitude, and amplify towards the end of June and beginning of July.

We next determine the Northern Hemisphere summer solstitial climatology of the horizontal winds by averaging over all measurements between 8 June and 7 July shown in Figure \ref{fig:u_alt_lat_days_junjul}. The results are plotted in \Figure \ref{fig:f7_u_v_alt_binned_junjul20} in the form of altitude-latitude and altitude-local time distributions for both the zonal (upper panels) and meridional (lower panels) components. In the lower thermosphere, the eastward winds are up to 40 m~s$^{-1}$, and the daytime westward mean zonal wind in the upper thermosphere can exceed --60 m~s$^{-1}$, especially between $15^\circ$ and $40^\circ $N. The westward mean zonal winds decrease around 160 km, such that the jet exhibits a split in altitude, especially, between 10-40$^\circ$N. In the lower thermosphere, meridional winds are weakly southward (up to --20 m~s$^{-1}$) between $\sim$90--105 km and northward between $\sim$105--120 km. Above 120 km, the meridional winds are directed southward with speeds occasionally exceeding --60 m~s$^{-1}$, for example at low-latitudes of the Southern Hemisphere during daytime after dawn and before dusk.
The meridional winds around $20^\circ-40^\circ $N are, generally, slower than at low-latitudes. In the lower thermosphere, both zonal and meridional components exhibit a distinct local time variability, when all observed latitudes are considered. Overall, the observed monthly-mean daytime meridional winds are directed southward and nighttime winds are northward.

Another view of the observed winds is presented in \Figure \ref{fig:f8_u_v_latlt_junjul20}, where the latitude-local time cross-sections of the mean zonal (left) and meridional (right) winds are shown at three representative thermospheric altitudes. Within the 90--105 km layer, the zonal winds are mainly eastward at midlatitudes around dawn and dusk and vary semidiurnally. At equatorial latitudes, the winds are eastward in the morning sector and westward in the afternoon, suggesting a diurnal variation. Meridional winds are southward during day and northward at night, which is indicative of  a diurnal signal as well. Higher up in the lower thermosphere between 105--120 km, mean zonal winds exhibit a more complex latitude-local time variability, however, meridional winds overall maintain a diurnal behavior at low-latitudes. In the upper thermosphere, the zonal winds are strongly westward at midlatitudes around dawn exceeding 100 m s$^{-1}$, moderately westward in general during day, and reverse their direction to strong eastward flow before dusk. Again, the low airglow signal could have potentially affected the magnitude of these winds at the terminator. Southward meridional winds dominate in the upper thermosphere during day, similar to the lower thermosphere. They indicate the global north-to-south branch of the solstitial circulation cell in the upper atmosphere.

\Figure \ref{fig:f8_u_v_latlon_junjul20} presents the latitude-longitude cross-sections of the zonal and meridional winds at three representative altitudes in the thermosphere. It shows that the monthly-mean morphology of the horizontal winds, i.e., wind magnitudes and directions, significantly changes in the lower thermosphere between 90 and 120 km. Within the 90--105 km layer, eastward winds of up to 40 m~s$^{-1}$ and southward winds of up to --30 m~s$^{-1}$ are prevalent in the Northern Hemisphere. The Southern Hemisphere low-latitudes are characterized by relatively slow westward winds. Around 105--120 km altitude, zonal winds in the Northern Hemisphere reverse the direction to westward and the northward flow becomes more prevalent compared to that at 90--105 km. The upper thermosphere at 185--200 km is dominated by westward and southward winds with speeds exceeding --$60 $ m~s$^{-1}$ and --40 m~s$^{-1}$, respectively. The latter are a part of the pole-to-pole solstitial meridional circulation. These upper thermosphere winds are, generally, much faster and more homogeneous than in the lower thermosphere.

Atmospheric circulation associated with neutral winds play an important dynamical role in redistribution of momentum, energy, and mass in the thermosphere. The associated latitude-local time and latitude-longitude and distributions of the mean horizontal circulation are seen in terms of  velocity vectors in \Figures \ref{fig:f11_circulation_latlt_junjul20} and \ref{fig:f12_circulation_latlon_junjul20} at three representative altitudes in the thermosphere. This representation of the winds shows the direction as well as the flow speeds. Upward and downward directed vectors represent northward (towards the North Pole) and southward flow (towards the South Pole). While vectors directed to the right and left are for eastward and westward flow, respectively, in latitude-longitude cross-sections, they facilitate an interpretation of the winds relatively to the day-night sectors in latitude-local time plots. For example, winds flowing from the dayside to the nightside or vice versa would be revealed. A large degree of spatiotemporal variability is seen especially within the lower thermosphere. Eastward and southward winds prevail in the layer between 90-105 km and westward and northward as well as southward flows are found within 105--120 km. Between 90-105 km, dusk-to-dawn and nighttime poleward flow is followed by a daytime equatorward flow. The mean circulation between 105--120 km exhibits the greatest degree of complexity in terms of varying scale sizes and vortices, generally directed westward during day and eastward during night diverging around the subsolar point. Generally flow speeds are in the order of 50 m s$^{-1}$. In the upper thermosphere, the observed daytime circulation is more easily discernible with clear diverging patterns around the subsolar point. Geographically south-westward  circulation prevails with flow speeds exceeding 100 m s$^{-1 }$ at midlatitudes.

  \subsection{{Thermospheric Mean Winds and Circulation during Northern Hemisphere
    Winter Solstice}}

{So far we discussed the Northern Hemisphere summer season. Energy deposition from the Sun varies  seasonally changing the differential heating and the pressure gradient force, which alters the circulation patterns,  hence atmospheric wave propagation and its effects.}

{
  In order to complement the results presented above, we next analyze the winds and circulation patterns during the Northern Hemisphere winter. Figure \ref{fig:f13_u_v_alt_binned_dec19jan20} shows the  altitude-latitude and altitude-local time distributions of the mean zonal and  meridional winds averaged from 16 December 2019--31 January 2020, which is representative of the December solstice (or Northern winter) conditions. {They are} presented in a manner similar to June solstice mean winds (Figure
  \ref{fig:f7_u_v_alt_binned_junjul20}). Note that this period does not include any  stratospheric sudden warming (SSW) events \citep{roy_2022_DynamicalEvolutionSudden}. It is seen that the zonal winds are 
  westward ({with magnitudes $<$} 40 m s$^{-1}$) at low-latitudes and eastward at midlatitudes ( {with magnitudes $<$} 20 m s$^{-1}$), relatively slower than {the} summer winds. During winter, slow  meridional winds are seen between 90--120 km, and strong eastward daytime winds prevail above 120 km. Local time variations in the lower thermosphere indicate a semidiurnal variation of the zonal winds and a diurnal variation of the meridional winds in the lower thermosphere, which are qualitatively somewhat similar to summer  conditions, however, the wintertime daytime northward and nighttime southward phase of the meridional wind variations are different compared to the summer one. In the upper thermosphere, daytime meridional winds are northward exceeding 60 m~s$^{-1}$  at midlatitudes and flow easterly at dawn, and somewhat westerly and easterly at dusk.
}

{
  Figure \ref{fig:f14_circulation_latlon_dec19jan20} shows the associated horizontal  circulation during Northern winter season, as was done for the summer solstice (Figure  \ref{fig:f12_circulation_latlon_junjul20}). In the lower thermosphere, the circulation is overall easterly and southward (winter-to-summer), with geographically varying intensity. The upper thermosphere during daytime is  dominated by a westward and northward (summer-to-winter) flow, becoming dominated by meridional component at midlatitudes. 
}

\section{Discussion}
\label{sec:discussion}
We have presented the mean zonal and meridional winds on 20 June 2020 as well as
averaged over a one-month {summer solstice} period (8 June--7 July 2020)
{and 47-day winter solstice period (16 December 2019--31 January 2020)}, using
continuous measurements of ICON/MIGHTI. The climatology of {thermospheric} horizontal circulation for the summer and winter solstice period{s} have been constructed for the first time. We next discuss dynamical forces influencing the winds and some of the noteworthy features of the observations, comparing our analysis to previous studies.

\subsection{Dynamical Forces that Control Upper Atmospheric Winds}
Multiple observations demonstrate that winds are extremely variable, especially in the lower thermosphere \citep[{e.g.,}][]{Larsen02, Larsenfesen09, Lehrmacher_etal22}. In order to characterize their climatology, the data have to be averaged using appropriate bins over multiple days. What processes drive the mean and variable structure of the winds is of great interest.  For this, it is instructive to discuss first the dynamical forces that control the motion of an air parcel. The complete momentum balance for neutral winds is given by \citep[][p. 112]{Yigit18}
\begin{linenomath*}
\begin{equation}
  \label{eq:mom_balance}
  \frac{\partial\mathbf{u}} {\partial t} =
  - (\mathbf{u}\cdot\del) \, \mathbf{u}
  - \frac{1}{\rho} \del P
  + \mathbf{g}
  - 2 \mathbf{\Omega} \times \mathbf{u}
  + \frac{1}{\rho} \del\cdot {\bm\tau}
  - \sum_k \nu_{nk} (\mathbf{u} - \mathbf{u}_k)
  + \mathbf{f}^\prime, 
\end{equation}
\end{linenomath*}
where $\mathbf{u} = (u,v,w) $ is the neutral wind vector, $P = \rho R T/M$ is the thermodynamic pressure, with temperature $T$, mass density $\rho $, universal gas constant $R $, and molar mass $M$; $\mathbf{g}$ is gravitational acceleration, $\mathbf{\Omega} $ is the rotation rate of Earth, ${\bm\tau} $ is the viscous shear stress, $\mathbf{u}_k$ is the velocity of particles of species $k$, which the neutrals with collision frequencies $\nu_{nk}$ collide with, and $\mathbf{f}^\prime $ is the momentum deposition  by eddies or small-scale waves. On the right hand side of (\ref{eq:mom_balance}) from left to right the momentum balance terms are advective forcing, pressure gradient force, gravity, Coriolis force,  viscous stress, frictional drag due to interactions of neutrals with charged particles, e.g., ion drag, and wave-induced momentum deposition. For an incompressible atmosphere the viscous shear stress is proportional to the vector laplacian of the wind velocity, i.e., $ \rho^{-1} \del\cdot {\bm \tau} = \nu \laplacian \mathbf{u}$, where $\nu=\mu/\rho $ is the kinematic viscosity and $\mu$ is the dynamic viscosity. In numerical models, the wave-induced momentum  forcing $\mathbf{f}^\prime $ is often not resolved  and has to be parameterized \citep[e.g.,][]{Yigit_etal08, MedvedevYigit19}. With improving resolution, global numerical models are increasingly able to capture a larger portion of subgrid-scale wave effects \citep[e.g.,][]{Miyoshi_etal14}. Depending on the altitude, latitude, local time, and time scales of motion, various combinations of these dynamical forces shape the neutral wind circulation. In general, atmospheric dynamics is nonlinear, which can lead to winds and circulation over a broad spectrum of spatiotemporal scales and complexity, as demonstrated by ICON/MIGHTI wind analysis presented earlier. 

\subsection{Thermospheric Winds and Circulation during Solstice}
\label{sec:horiontal-winds-at}
\begin{figure}[t]
  \includegraphics[width=1.0\textwidth,  keepaspectratio, height=0.9\textheight]{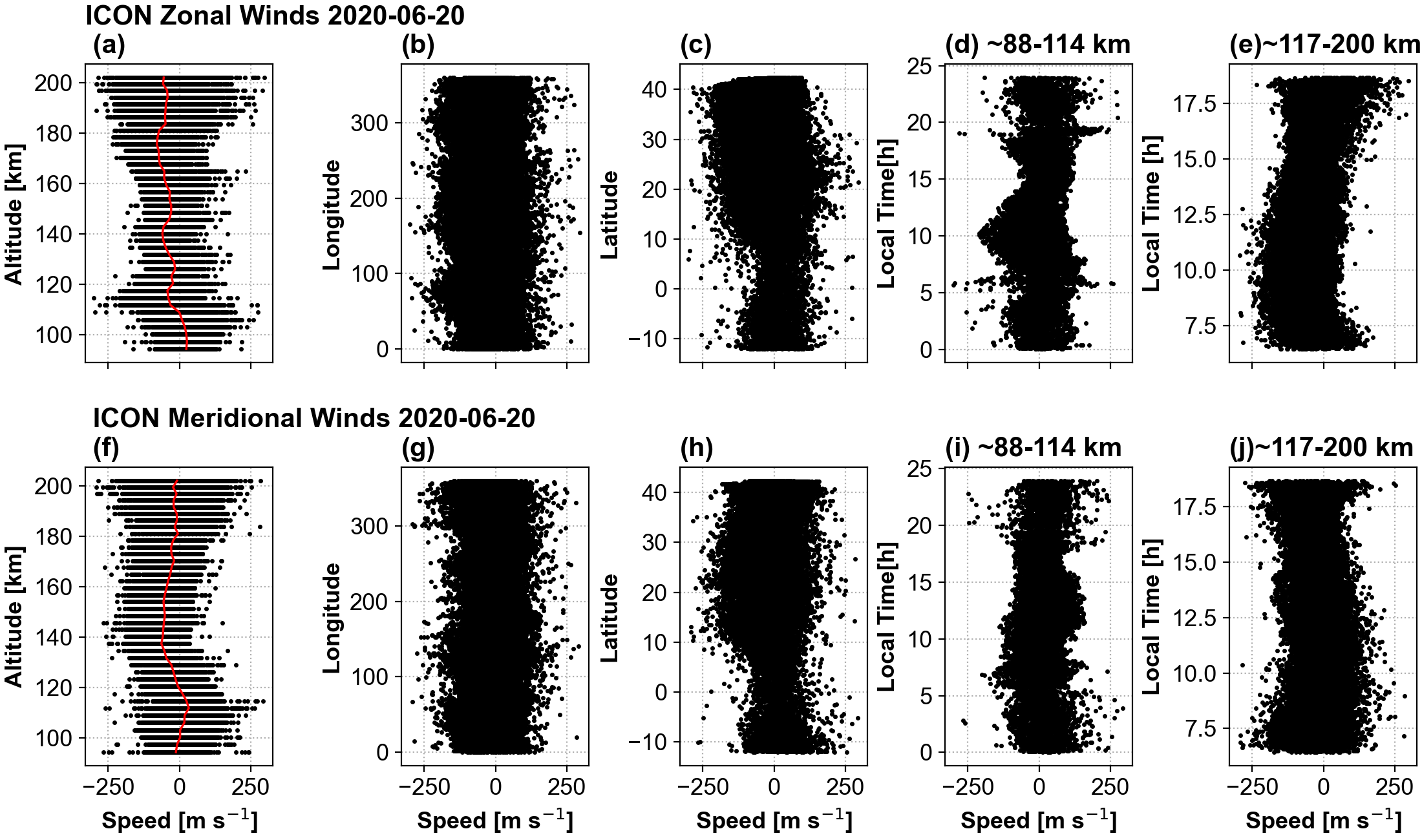}
  \caption{Altitude, longitude, latitude, and local time variations of the zonal (upper panels) and meridional (lower panels) winds in  m s$^{-1}$ during 20 June 2020 (i.e., Northern Hemisphere Summer solstice) as observed by ICON/MIGHTI. Longitude and latitude variations include data from all altitudes between 88-200 km, while the local time variations of the winds are shown for the lower thermosphere 88-{114}  km and upper thermosphere separately {117} - 200 km. Mean winds are shown with the red curve. See Figure \ref{fig:ICON_coverage} for the details of the spatiotemporal coverage.}
  \label{fig:all_winds}
  \end{figure}

\begin{figure}[t]
    \includegraphics[width=0.8\textwidth, keepaspectratio, height=0.92\textheight]{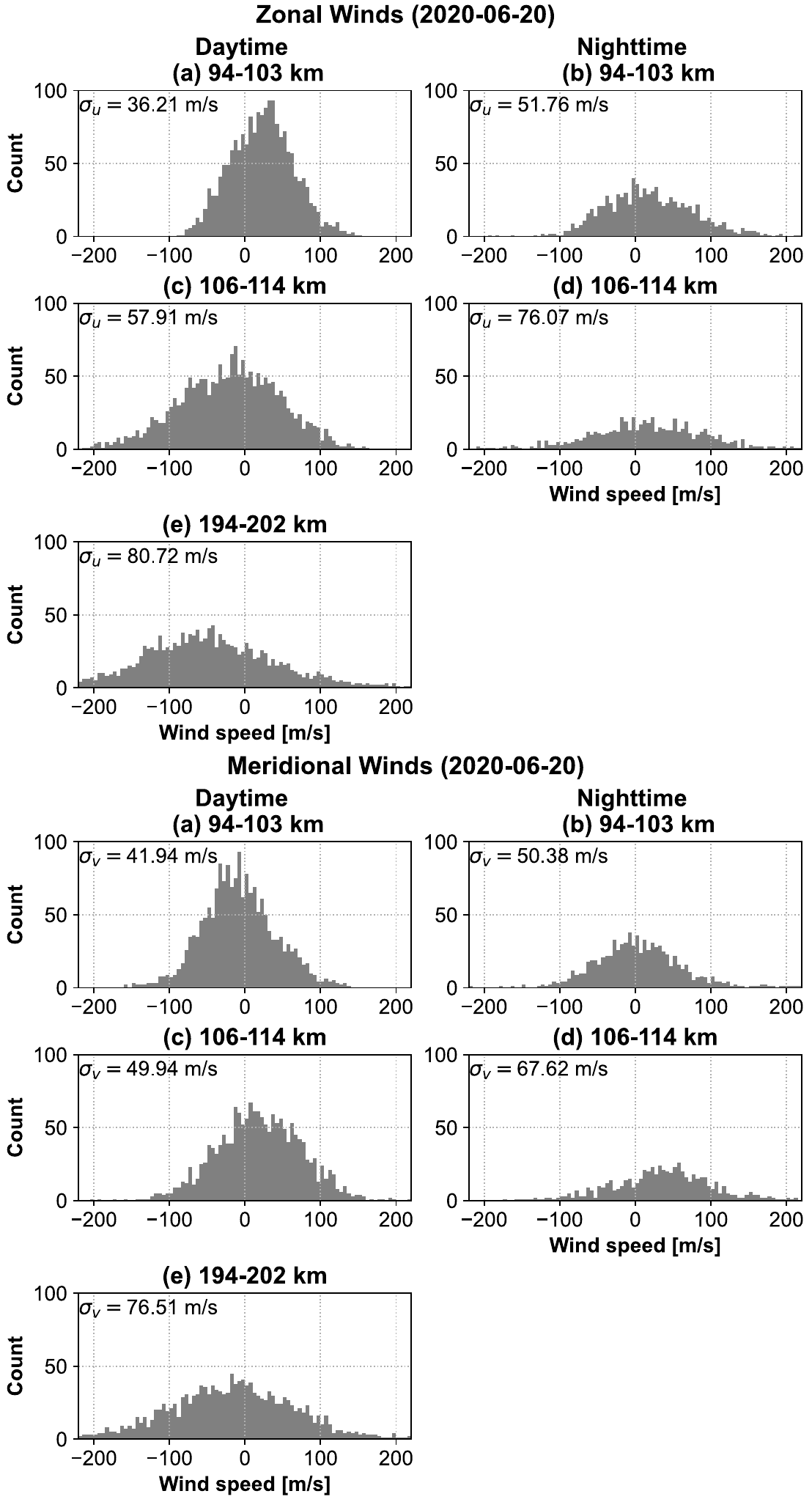}
    \caption{Distributions of the zonal and meridional wind velocities by day and night on 20 June 2020. Wind speeds are binned in 5 m s$^{-1}$ intervals at three representative altitudes in the thermosphere. Each altitude layer includes equal number of data points. Standard deviations are given in each plot in the upper left corner.} 
    \label{fig:variability}
  \end{figure}

In order to illustrate one possible way of studying wind variability seen in \Figure \ref{fig:all_winds}, we have plotted the zonal and meridional wind speeds as a function of their occurrence rates on 20 June 2020 (\Figure \ref{fig:variability}.) They generally exhibit a Gaussian distribution centered around slow speeds. Larger standard deviation found at night suggests that winds are more variable during night, which, in turn, indicates an elevated atmospheric wave activity. This is also suggested by more variable wind vectors at night (\Figure \ref{fig:f11_circulation_latlt_junjul20}). The atmosphere is in general less dissipative at night due, for example, to smaller molecular viscosity, providing more favorable propagation conditions for small-scale waves. The wind variability is found to increase with height, which could be linked to larger wind speeds at higher altitudes and growing with height wave amplitudes.

\begin{figure}[t]
  \includegraphics[width=1.05\textwidth,  keepaspectratio, height=0.9\textheight]{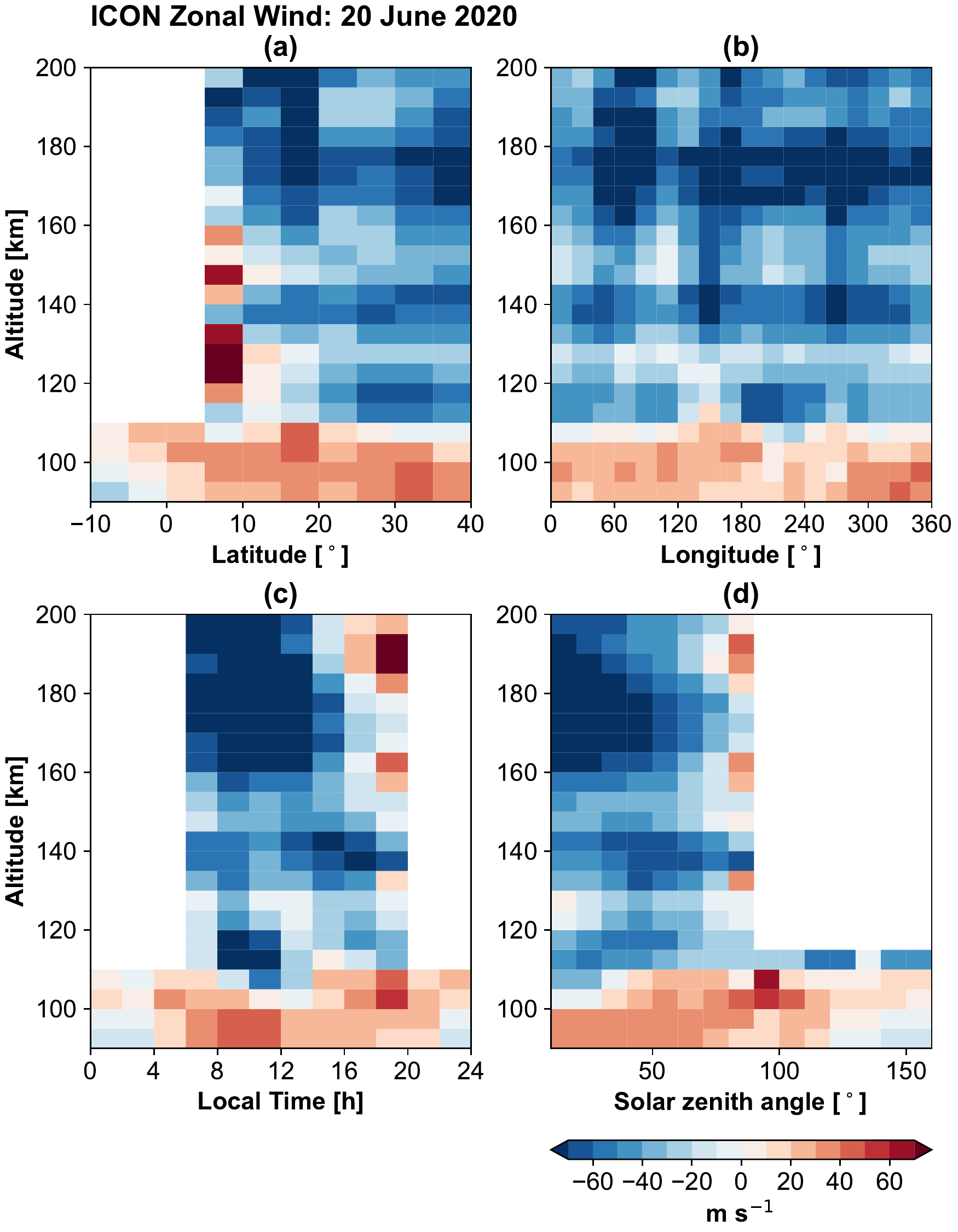}
  \caption{Altitude distributions of zonal wind speed (m s$^{-1}$)  as a function of latitude, longitute, local time, and solar zenith angle as observed by ICON/MIGHTI on 20 June 2020.}
  \label{fig:zonal_winds_binned}
\end{figure}

If dissipative effects such as wave breaking/saturation and viscosity are ignored in (\ref{eq:mom_balance}), then the large-scale behavior of the thermospheric mean winds is controlled primarily by pressure gradient force generated by the differential heating by the Sun and, to a secondary degree, by inertia (advection), ion drag, and Coriolis force, with the latter being negligible at equatorial latitudes. This simplified force balance should yield westward (eastward) mean zonal winds in the summer (winter) hemisphere and summer to winter mean meridional flow, with associated upwelling in the summer hemisphere and downwelling in the winter hemisphere as a consequence of mass continuity \citep{Forbes07}. However, as observed by ICON/MIGHTI,  solstitial zonal mean winds are consistently  eastward in the upper mesosphere and lower thermosphere (MLT) between 90--110 km (\Figures \ref{fig:zonal_winds_binned}, \ref{fig:f7_u_v_alt_binned_junjul20}) with increasing magnitude from low- to middle-latitudes. This feature of the zonal winds is associated with the eastward gravity wave momentum deposition in the MLT, as has been demonstrated in a number of general circulation modeling studies \citep{Garcia_etal07, Yigit_etal09, MiyoshiYigit19, Lilienthal_etal20, Griffith_etal21}. Eastward mean winds of up to 10--40 m~s$^{-1}$ were also seen around 10-40$^\circ$N in the monthly mean wind climatologies compiled as part of the Upper Atmosphere Research Satellite Reference Atmosphere Project UARS \citep{SwinbankOrtland03} and at northern midlatitudes in meteor and MF radar measurements \citep{Conte_etal17, Tang_etal21, Portnyagin00}. They are a general feature of the summer midlatitudes in the MLT \citep{Drob_etal08, Smith12}. The observed dawn-dusk asymmetry in the lower thermospheric winds is noticeable as well (\Figure \ref{fig:zonal_winds_binned}c).

Lower thermospheric winds have been routinely observed by incoherent scatter radars (ISR) and meteor radars. \citet{Portnyagin_etal99}'s analysis of seasonal variations of the mean zonal wind observed by ground-based radars and WINDII at 95 km shows eastward winds of 40--50 m~s$^{-1}$ at midlatitudes for Northern Hemisphere summer conditions, which is similar to ICON/MIGHTI measurements. \citet{Zhang_etal03} used the Millstone Hill incoherent scatter radar (42.6$^\circ$N) to study the seasonal climatology of zonal and meridional winds in the ionospheric $E$-region (94--130 km) and compared with WINDII observations. These results shown in their figures 2 and 3 demonstrate semidiurnal variations during all seasons. Our monthly mean zonal and meridional winds during the solstice viewed as altitude-local time cross-sections indicate a semidiurnal variation as well (\Figure \ref{fig:f7_u_v_alt_binned_junjul20}). Although the Millstone ISR provides data at a fixed latitude and here we have included all latitudes between 10$^\circ $S-40$^\circ $N, the phases of the semidiurnal variations in mean winds are quite similar. Overall,  differences in the magnitudes are probably due to differences in latitude and seasonal binning.

ICON/MIGHTI provides evidence that the mean meridional winds in the MLT reverse their direction. The northward mean meridional flow (i.e., from winter to summer) seen in the MLT is opposite to the radiatively driven mean meridional flow, which is due to the direct response of the circulation to the eastward gravity wave momentum deposition in the MLT \citep{Holton83}. A simple force balance based on the Transformed Eulerian Mean (TEM) analysis illustrates the role of the zonal wave forcing in driving the mean residual circulation at midlatitudes:
\begin{linenomath*}
  \begin{equation}
  \label{eq:momentum-eddy}
  \frac{\partial \bar{u}} {\partial t} - f_0 \bar{v}^* = \bar{a}_x,
\end{equation}
\end{linenomath*}
where $\bar{a}_x$ is the total zonal force due to small-scale and  non-zonal eddies/waves, $f_0$ is the Coriolis parameter at midlatitudes, and $\bar{v}^*$ is the meridional component of the residual circulation ($\bar{v}^*, \bar{w}^*$), where $\bar{w}^*$ is the vertical component, and the overbar indicates an appropriate averaging as seen in (\ref{eq:averaging}). For steady-state conditions, (\ref{eq:momentum-eddy}) implies that the mean meridional circulation (or transport) is primarily driven by wave dissipation, i.e., $\bar{v}^* \approx - \bar{a}_x / f_0 $. Around the midlatitutude MLT, small-scale gravity waves are the primary contributor to the zonal body force. Diurnal migrating tide-gravity wave interactions are an important mechanism of wind variability, especially in the low-latitude MLT \citep{MiyaharaForbes91, WatanabeMiyahara09, YigitMedvedev17}.

\begin{figure}[t]
  \includegraphics[width=1.05\textwidth,  keepaspectratio, height=0.9\textheight]{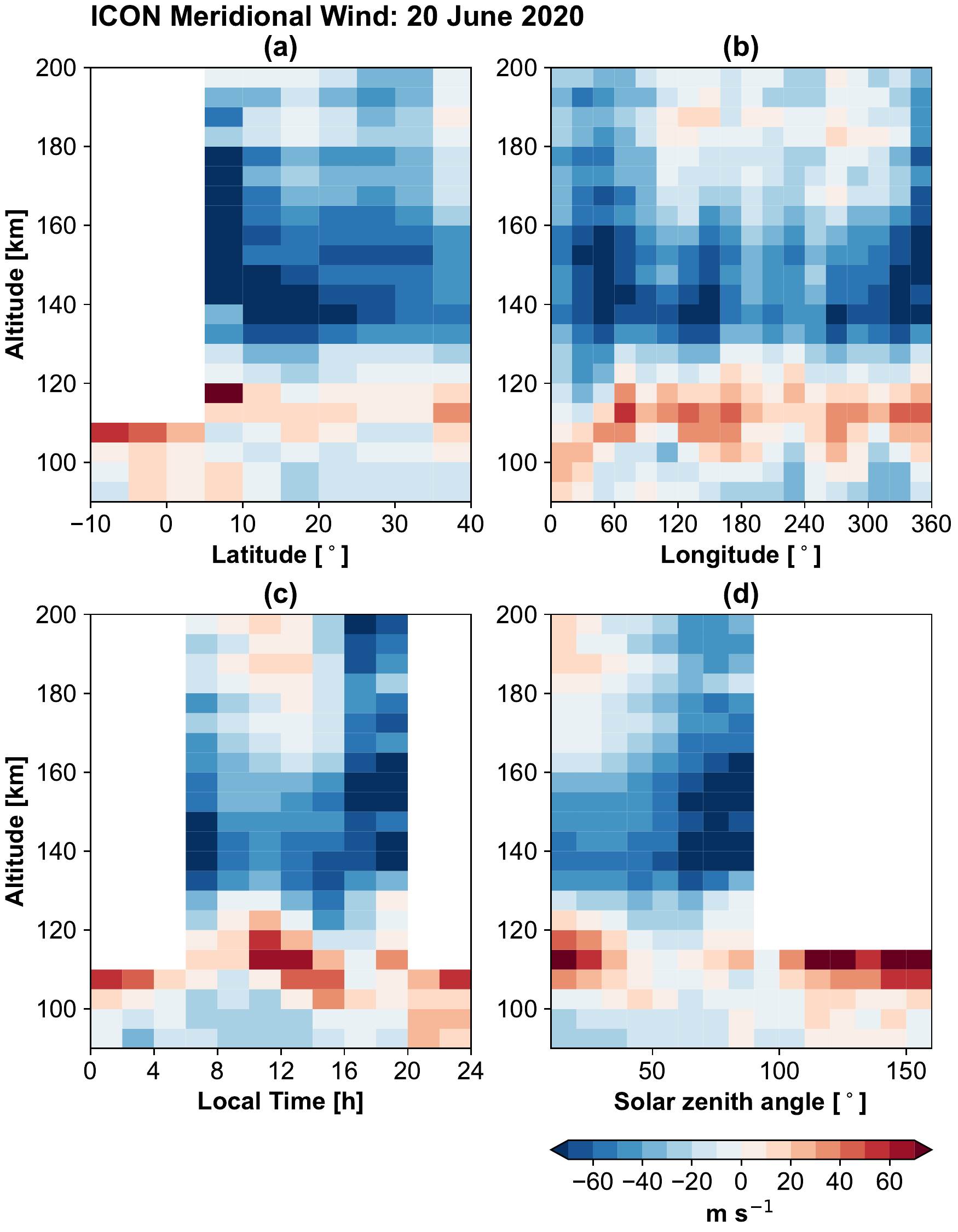}
  \caption{Same as Figure \ref{fig:zonal_winds_binned}, but for the meridional winds.}
  \label{fig:meridional_winds_binned}
  \end{figure}

\begin{figure}[t]
  \includegraphics[width=1.1\textwidth,  keepaspectratio, height=0.9\textheight]{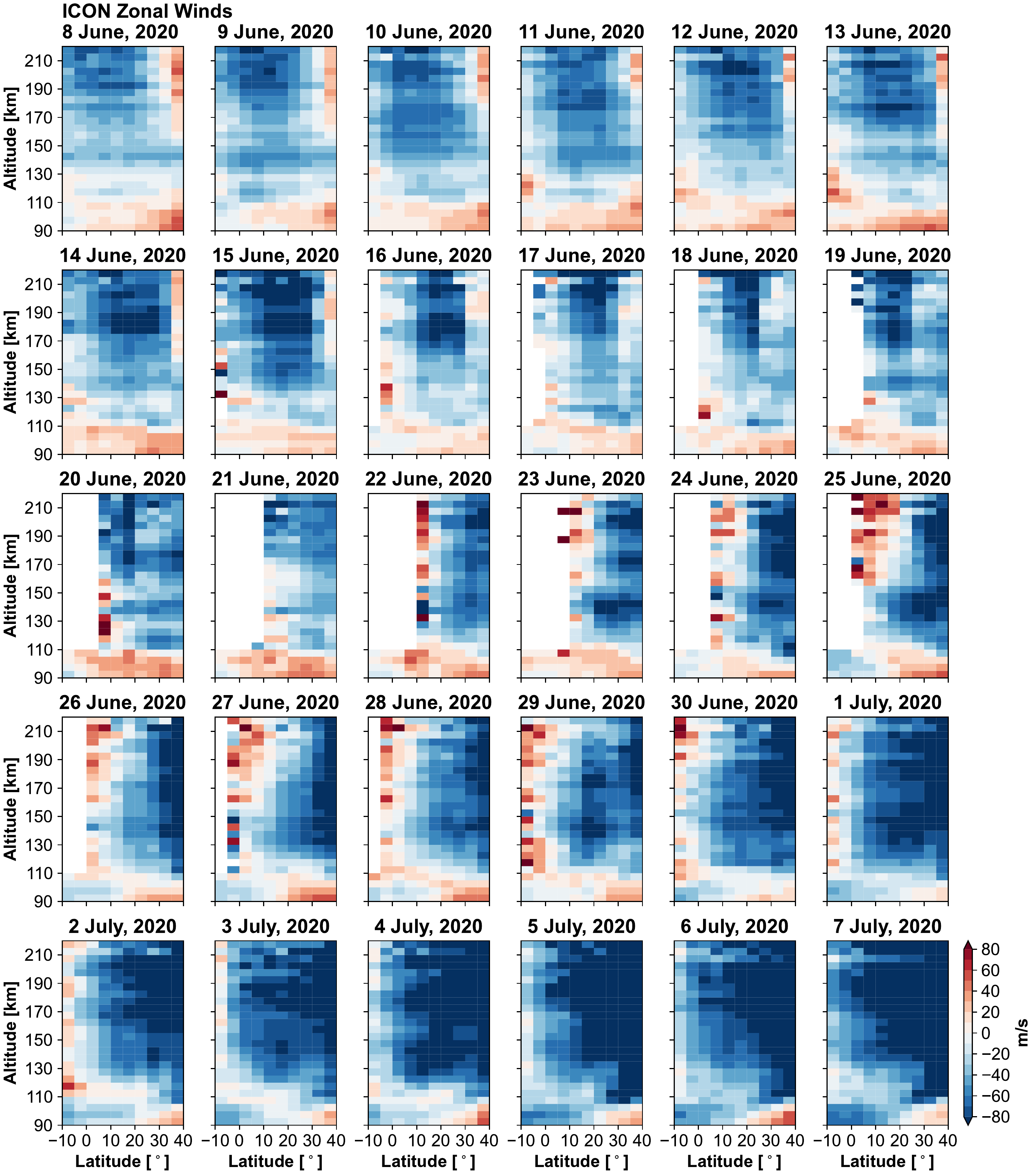}
  \caption{Day-to-day variations of the altitude-latitude distributions of the thermospheric zonal winds between 90-200 km and $10^\circ$S - $40^\circ$N in m s$^{-1}$ as observed by ICON/MIGHTI from 8 June - 7 July 2020. For each day, all observed longitudes and local times/solar zenith angles are included. }
  \label{fig:u_alt_lat_days_junjul}
  \end{figure}

\begin{figure}[t]
    \includegraphics[width=1\textwidth,
    keepaspectratio, height=0.9\textheight]{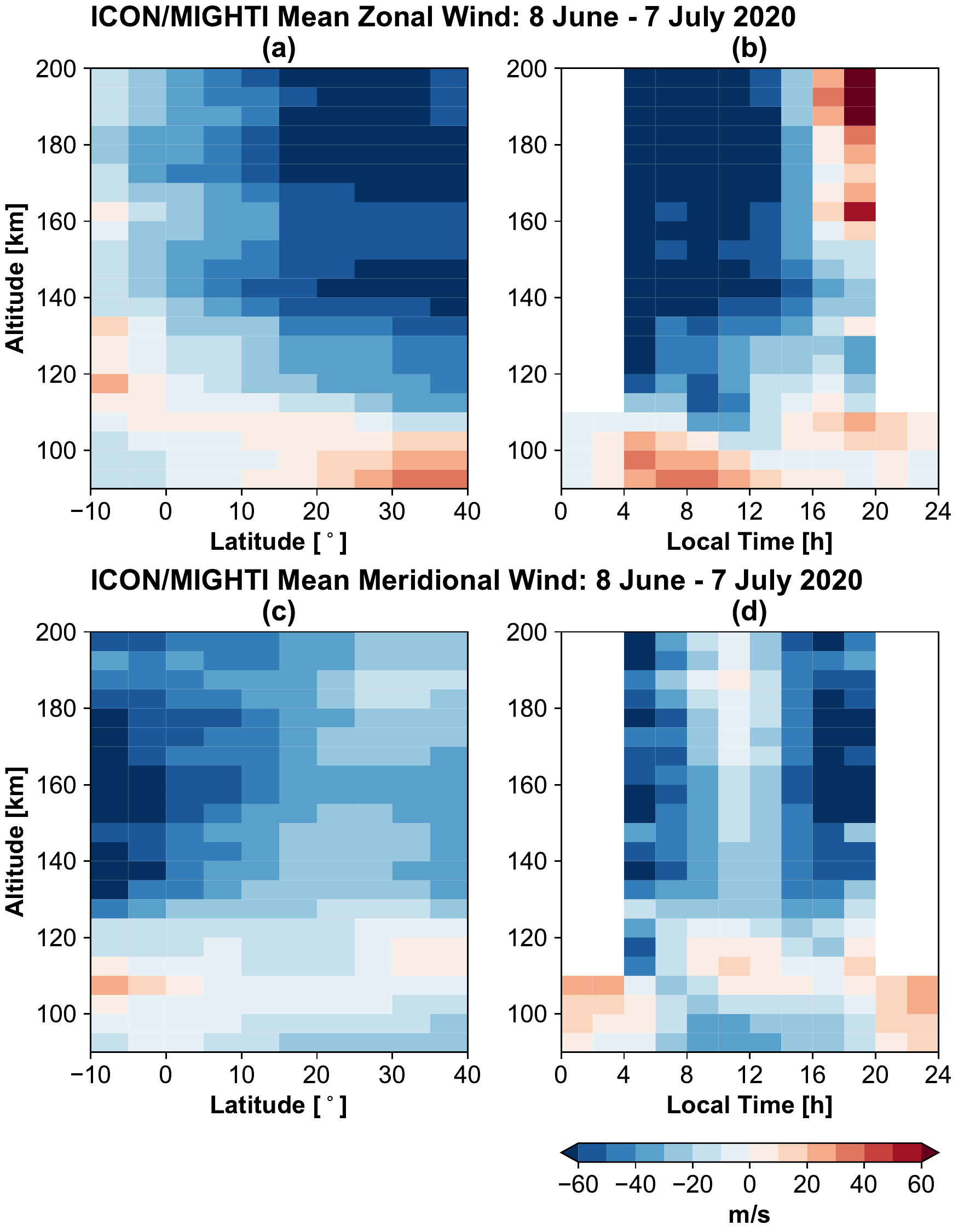}
    \caption{Zonal and meridional wind climatology from 90 to 200 km presented as altitude- latitude and altitude-local time cross-sections based on ICON/MIGHTI data from 8 June -- 7 July 2020. Data include daytime and nighttime measurements below 110 km and only daytime observations above 110 km. }
    \label{fig:f7_u_v_alt_binned_junjul20}
  \end{figure}

\begin{figure}[t]
    \includegraphics[width=1\textwidth,
    keepaspectratio, height=0.9\textheight]{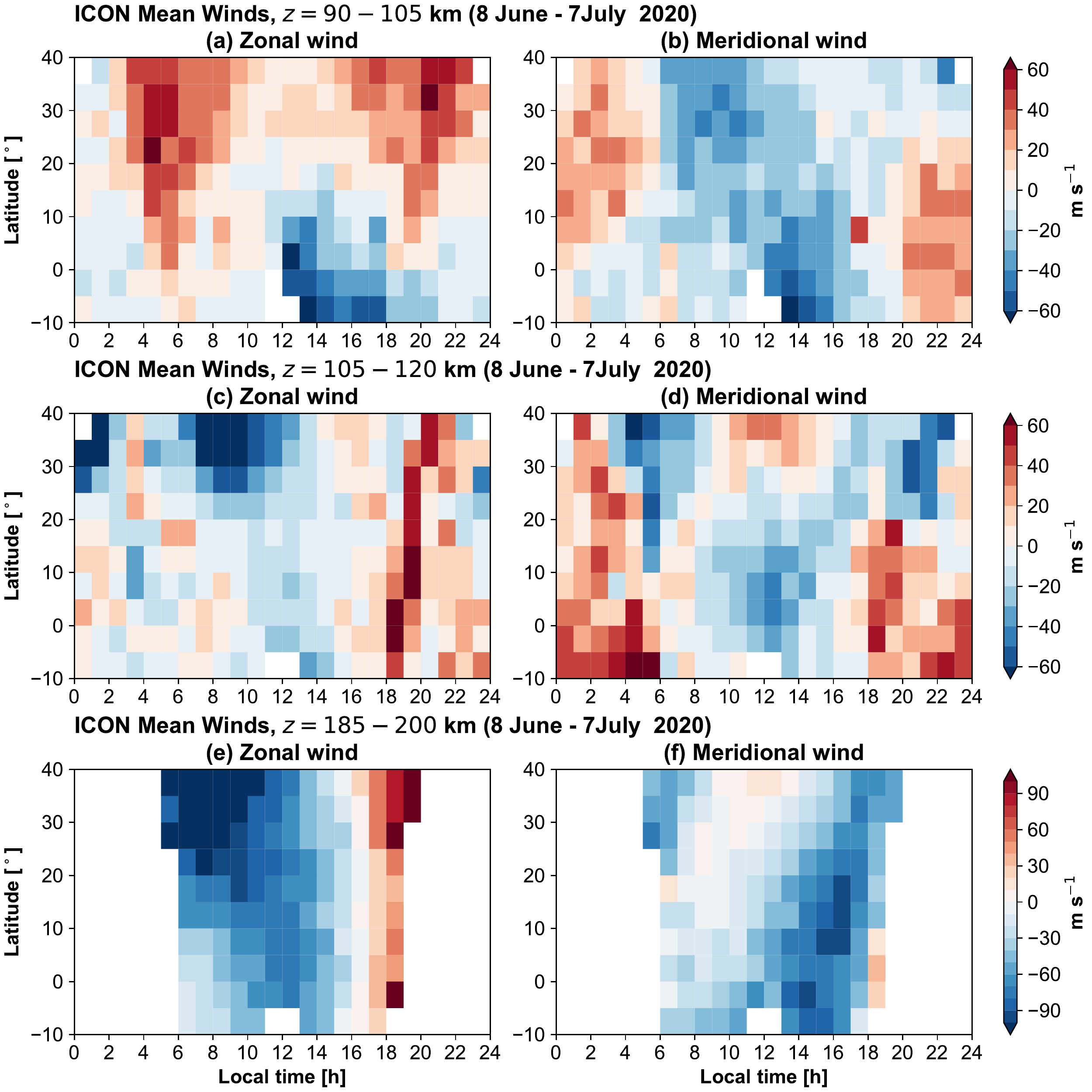}
    \caption{Latitude-local time distribution of zonal and meridional winds at three representative thermospheric altitudes. }    \label{fig:f8_u_v_latlt_junjul20}
  \end{figure}

\begin{figure}[t]
    \includegraphics[width=1\textwidth,
    keepaspectratio, height=0.9\textheight]{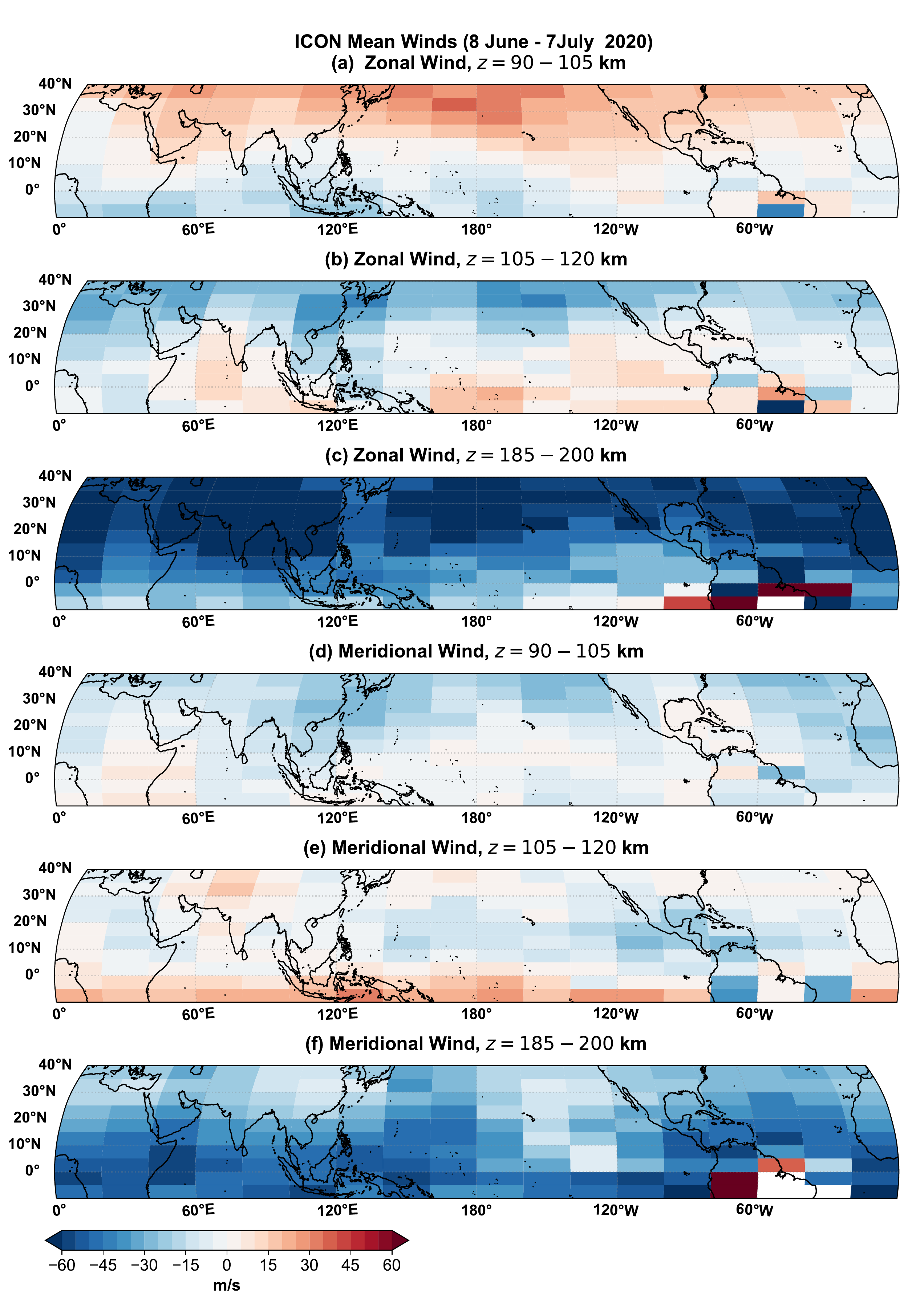}
    \caption{Latitude-longitude distributions of the zonal and meridional winds at three representative thermospheric altitudes: (a)-(b) 90-105  km, (c)-(d) 105-120 km, (e)-(f) 185-200 km. Note that winds at 185-200 km altitude are only daytime measurements.}    \label{fig:f8_u_v_latlon_junjul20}
  \end{figure}

\begin{figure}[t]
    \includegraphics[width=0.7\textwidth,
    keepaspectratio, height=0.9\textheight]{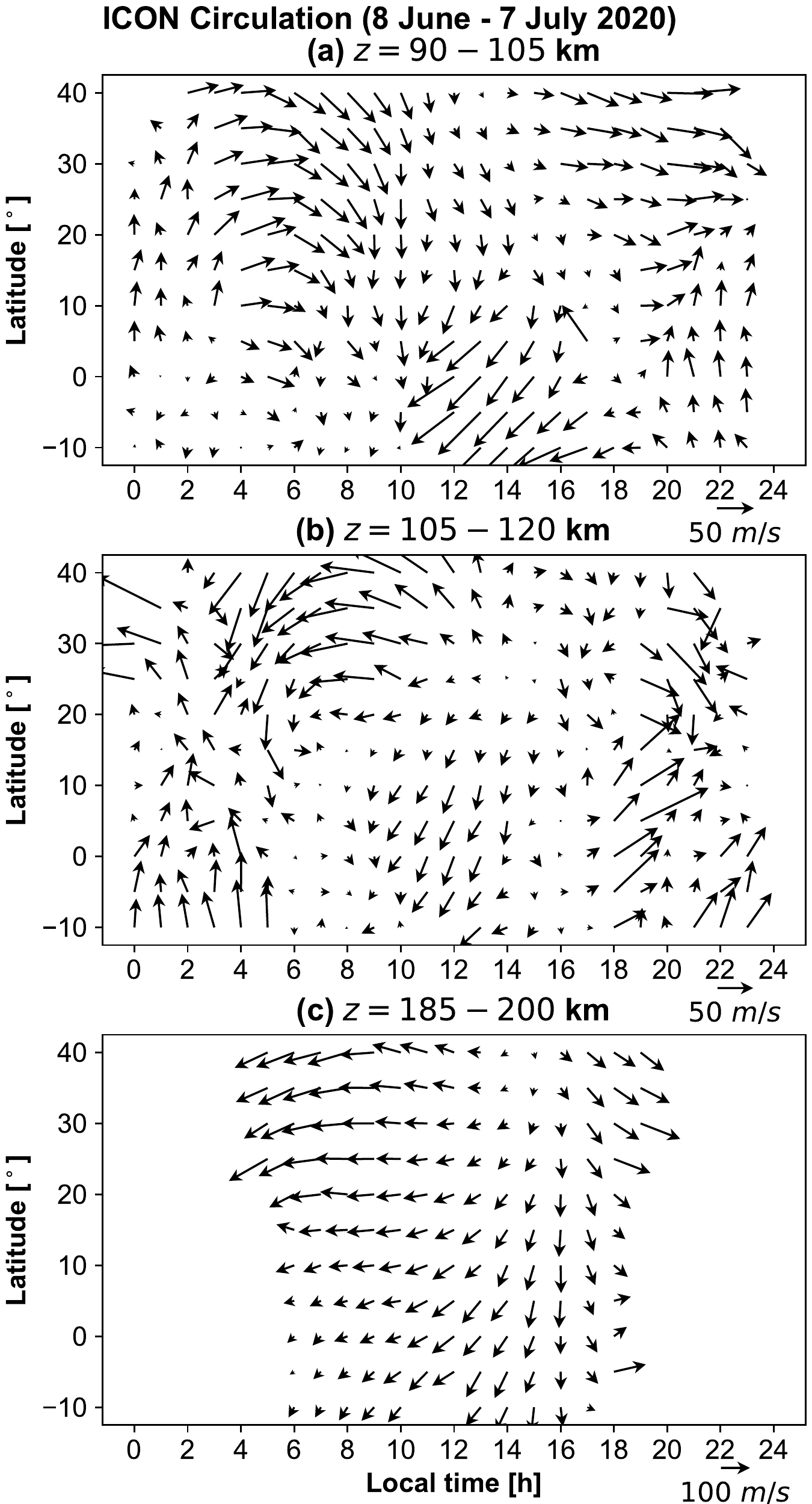}
    \caption{Latitude-local time distribution of the mean horizontal circulation
       at three representative thermospheric altitudes averaged over longitudes for 8 June--7July 2020. Note that the magnitude of the vector is 50 m s$^{-1}$ in panels a and b, while it is 100 m s$^{-1}$ in panel c.}
    \label{fig:f11_circulation_latlt_junjul20}
  \end{figure}

\begin{figure}[t]
      \includegraphics[width=1\textwidth,
      keepaspectratio, height=0.9\textheight]{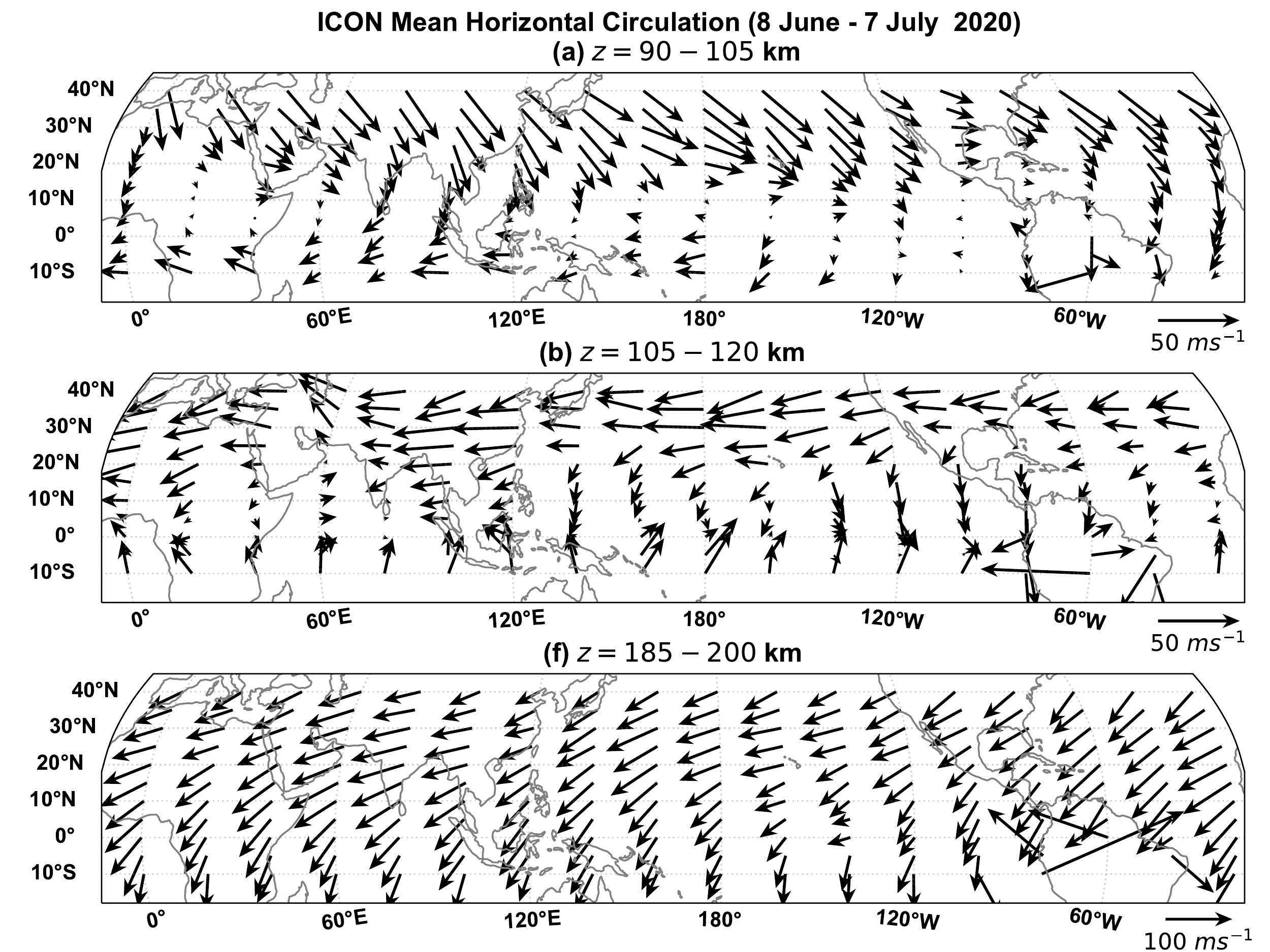}
      \caption{Same as Figure \ref{fig:f11_circulation_latlt_junjul20} but for the latitude-longitude distributions of the mean horizontal circulation
        at three representative thermospheric altitudes. Note that winds at 185-200 km altitude are only daytime measurements.}
      \label{fig:f12_circulation_latlon_junjul20}
    \end{figure}

Above 120 km, radiative processes and ion-neutral coupling (i.e., ion drag), which is proportional to the relative velocity of neutrals and ions moving within the magnetic field, become increasingly important in driving the horizontal circulation. ICON/MIGHTI observations show that the zonal winds are predominantly westward and southward (i.e., directed from summer to winter hemisphere) in the upper thermosphere (\Figures \ref{fig:zonal_winds_binned}--\ref{fig:f12_circulation_latlon_junjul20}). This large-scale flow is maintained by the pressure gradient force, modulated by ion drag and Coriolis effect. Molecular viscosity, which increases exponentially with height, smooths out smaller-scale motions, as can be seen in daytime circulation and winds between 185--200 km (\Figures \ref{fig:f11_circulation_latlt_junjul20}--\ref{fig:f12_circulation_latlon_junjul20}).

Recently, \cite{Drob_etal15} have updated the Horizontal Wind Model (HWM14) and provided a comparison against WINDII climatologies of \citet{Emmert_etal02} and an older version of HWM. Our altitude-local time analysis of mean winds at 200 km (\Figures \ref{fig:f8_u_v_latlt_junjul20}e-f) can qualitatively be compared with their June solstice climatology (day 180) at 250 km. There is a general agreement concerning the morphology of the mean winds. Westward zonal winds prevail during day and eastward winds around dusk at low- to midlatitudes in the Northern Hemisphere. Meridional winds are generally northward during morning and southward during afternoon.

\begin{figure}[t]
      \includegraphics[width=1\textwidth,
      keepaspectratio, height=0.9\textheight]{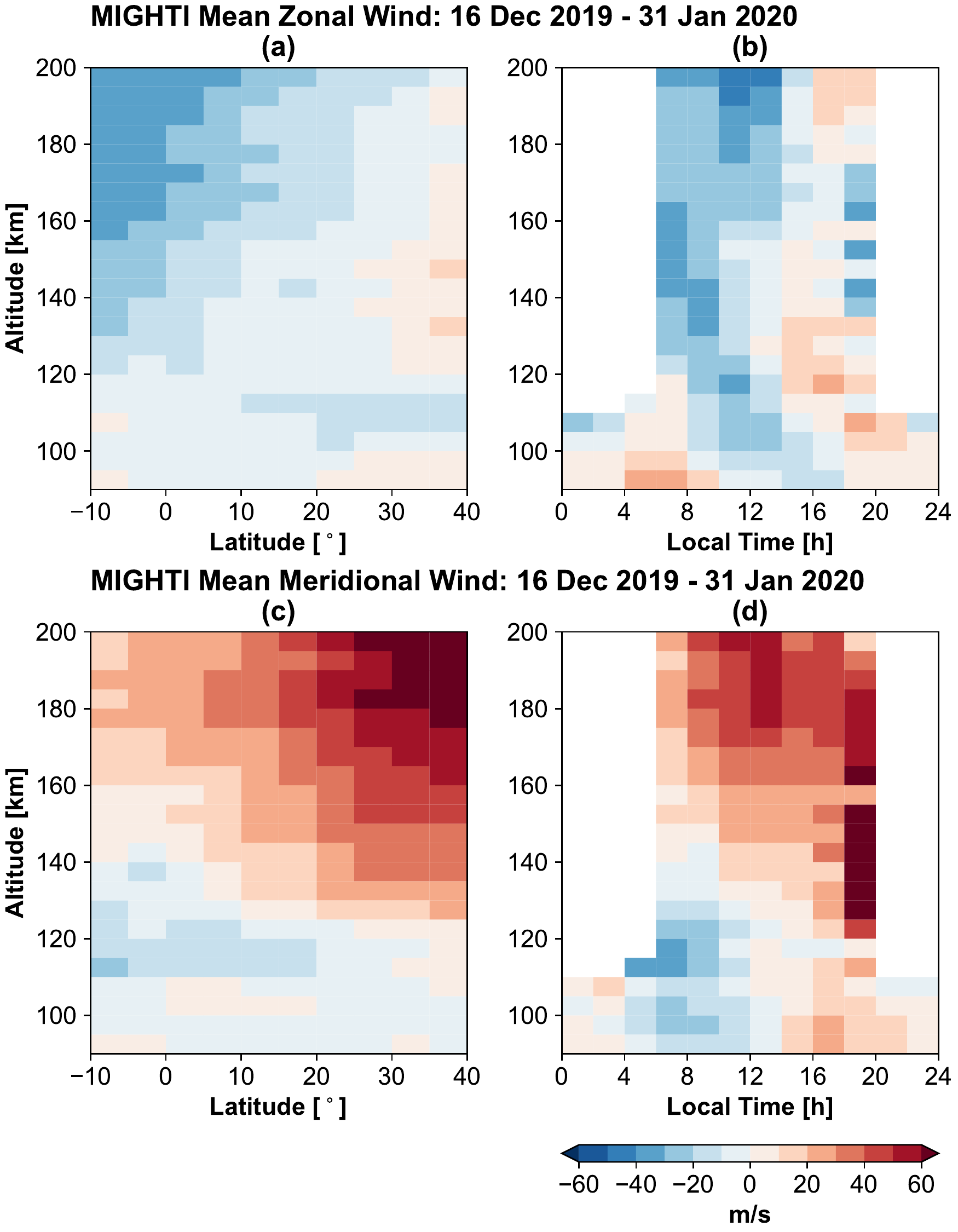}
      \caption{{Same as Figure \ref{fig:f7_u_v_alt_binned_junjul20} but for the
        Northern Hemisphere winter season based on ICON/MIGHTI observations from 16
        December 2019 -- 31 January 2020.}}
      \label{fig:f13_u_v_alt_binned_dec19jan20}
    \end{figure}

\begin{figure}[t]
      \includegraphics[width=1\textwidth,  keepaspectratio, height=0.9\textheight]{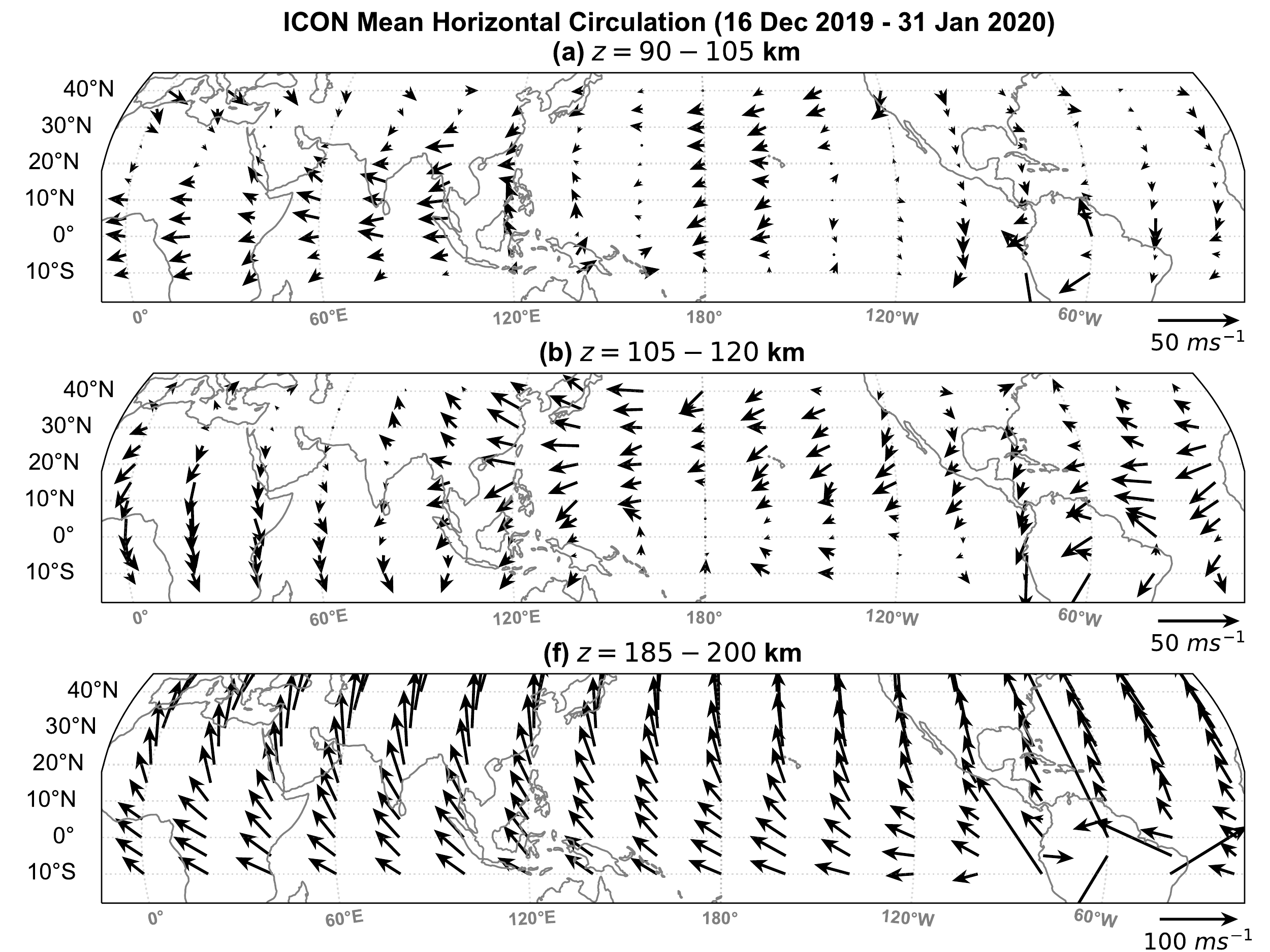}
      \caption{{Same as Figure \ref{fig:f12_circulation_latlon_junjul20} but for the
        Northern Hemisphere winter season based on ICON/MIGHTI observations from 16
        December 2019 -- 31 January 2020.}}
      \label{fig:f14_circulation_latlon_dec19jan20}
    \end{figure}

{In order to complement the northern summer solstice results, we have also compiled northern winter winds and circulation, including data from 16 December 2019 -- 31 January 2020, during which no SSWs
occurred \citep{roy_2022_DynamicalEvolutionSudden}. Although the data sampling is not the same between these seasons, the results can be at least qualitatively compared. There is certainly a certain degree of asymmetry between the summer and winter seasons in the Northern Hemisphere in terms of the magnitude and spatial distribution of winds and circulation. The mean zonal winds are much weaker during winter than summer. The lower thermospheric westward wind reversal (Figure
\ref{fig:f13_u_v_alt_binned_dec19jan20}a) is also weaker in magnitude than the eastward reversal during summer (Figure \ref{fig:f7_u_v_alt_binned_junjul20}), which qualitatively {agrees} with previous wind compilations in this region.}

{The summer-winter differences  in circulation can be caused  {by} a
  combination of seasonal changes in pressure forces, upward propagation condition{s for} gravity waves,
tides, and planetary waves. Since dynamical conditions {(westerly winds) are favorable} for the upward propagation of
planetary waves  in the winter hemisphere {and not in} the summer hemisphere with easterlies \citep{YigitMedvedev16}, presumably planetary waves contribute significantly to the differences between the two seasons.}

\section{Summary and Conclusions}
\label{sec:summary-conclusions}

We have presented the mean behavior of the thermospheric zonal and meridional winds
at 90--200 km as observed by the MIGHTI instrument onboard NASA's ICON spacecraft
between 10$^\circ$S and 40$^\circ$N. A comprehensive picture of the mean zonal and
meridional winds and horizontal circulation has been shown for a single solstice day,
20 June 2020, and using a month of continuous observations from 8 June--7 July 2020,
representative of Northern Hemisphere solstice conditions{, and a 47-day
  Northern Hemisphere winter solstice period from 16 December 2019--31 January 2020,
  primarily focusing on the northern summer solstice season}. 

The main inferences of our analysis of ICON/MIGHTI {northern summer solstice} winds are as follows:
\begin{enumerate}
\item Altitude, longitude, latitude, and local time profiles of winds show that the typical instantaneous zonal and meridional winds during solstice can exceed $\pm$150 m~s$^{-1}$, with with magnitudes increasing with altitude.

\item We have evaluated the occurrence rates of the wind speeds observed by ICON/MIGHTI on 20 June 2020 at three representative altitude bins 94--103 km, 106--114 km, and 194--202 km, such that each altitude bin included equal number of data. The speeds as a function of number of measurements exhibit a Gaussian distribution centered around small values. The nighttime magnitudes of the wind are greater than during day and increase as a function of height. Larger standard deviation at night suggests more variability during night, which indicate more atmospheric wave activity.
  
\item Thermospheric mean winds are up to $\pm 80$ m~s$^{-1}$ and depend strongly on altitude, latitude, longitude, and local time. Local time variations of the mean winds exhibit diurnal and semidiurnal variations. 
  
\item Mean winds and circulation change substantially within the lower thermosphere (90--120 km). Eastward and southward flow between 90--105 km change to a northward and westward flow within 105--120 km. Upper thermospheric winds are generally characterized by a westward and southward (i.e., directed from the summer-to-winter) flow in the Northern Hemisphere with diverging flow from the post-noon sector at midlatitudes. The upper thermospheric wind system is more homogeneous compared to the lower thermospheric one, which exhibits spatial variations at smaller scales and vortex patterns, especially around 105--120 km.

\item The observed eastward mean flow and the northward (winter-to-summer) meridional flows in the lower thermosphere are a consequence of eastward gravity wave momentum forcing there. ICON/MIGHTI observations are capable of demonstrating vertical coupling induced by waves. These features are in a good agreement with previous observations and modeling.

\item {Despite the seasonal differences in the data sampling size due to data gap issues, we have  qualitatively compared the northern summer and winter
  winds. Zonal winds during winter are clearly slower than summer zonal
  winds. Overall, an asymmetry between the summer and winter
seasons in terms of the magnitude and spatial distribution of winds and circulation is evident, which is presumably due to the differences in planetary wave propagation and
interaction with the mean flow as well as to differences in the generation,
propagation, and filtering in gravity waves and tides between winter and summer seasons.}
  
\item In the upper thermosphere, the morphology of the ICON/MIGHTI mean winds is, in general, in a good agreement with previous wind climatologies  based on WINDII and HWM14.

\end{enumerate}

The mean wind and circulation patterns inferred in this study using ICON/MIGHTI measurements can be used to constrain and validate general circulation models or as an input for numerical wave models. They also serve for an indirect validation of parameterized subgrid-scale processes, which control the large-scale winds and circulation.  This work is expected to contribute towards filling in the observational gap with horizontal winds in the thermosphere. 

\acknowledgments
This work was  supported by NASA (Grant Number 80NSSC22K0016). ICON is supported by NASA’s Explorers Program through contracts NNG12FA45C and NNG12FA42I.

\section*{Data Availability Statement}
The horizontal wind data (version 4) used in this study are available at the ICON data center (\url{https://icon.ssl.berkeley.edu/Data}).


\newpage



\appendix
\section{Data quality}
\begin{figure}
    \includegraphics[width=0.9\textwidth,  keepaspectratio, height=0.9\textheight]{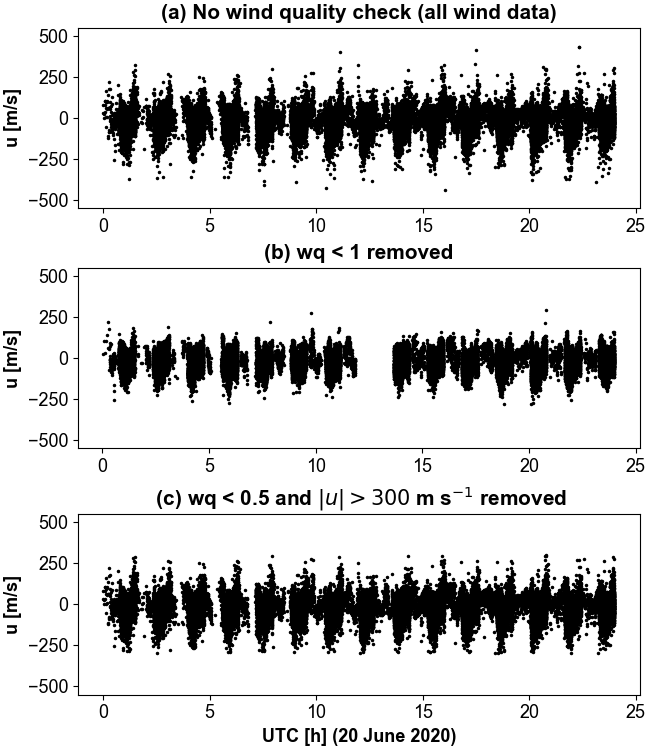}
    \caption{The effect of filtering the data according to quality is shown, where wq=1 stands for ``good", wq=0.5 is for ``good, but use with caution" and wq=0 is for ``bad" zonal wind measurements.}
    \label{fig:wind-quality-junjul2020}
  \end{figure}

We have removed data with quality less than 0.5 and filtered outliers by removing wind magnitudes exceeding 300 m s$^{-1}$. We illustrated the effect of this filtering for zonal winds in Figure \ref{fig:wind-quality-junjul2020}.

\end{document}